\documentclass[aps,prd,a4paper,onecolumn,amsmath,showpacs,superscriptaddress,nofootinbib,preprintnumbers,notitlepage]{revtex4-1}

\usepackage{graphicx}% Include figure files
\usepackage{dcolumn}
\usepackage{bm}
\usepackage{epsfig}
\usepackage{amsmath}
\usepackage{amsfonts}
\usepackage{amssymb}%
\usepackage{caption, subcaption, floatrow}
\begin{document}
\captionsetup{
  font=footnotesize,
  justification=raggedright,
  singlelinecheck=false
}
\captionsetup[figure]{labelfont=bf}
\title{Unifying   Inflation,  dark energy and dark matter with a scalar field and  exotic fermions}

\author{Eduardo Guendelman}
\email{guendel@bgu.ac.il}
\affiliation{Department of Physics, Ben-Gurion University of the Negev, Beer-Sheva, Israel.\\}
\affiliation{Frankfurt Institute for Advanced Studies (FIAS),
Ruth-Moufang-Strasse 1, 60438 Frankfurt am Main, Germany.\\}
\affiliation{Bahamas Advanced Study Institute and Conferences, 
4A Ocean Heights, Hill View Circle, Stella Maris, Long Island, The Bahamas.
}

\author{Ram\'{o}n Herrera}
\email{ramon.herrera@pucv.cl}
\affiliation{Instituto de F\'{\i}sica, Pontificia Universidad Cat\'{o}lica de Valpara\'{\i}so, Avenida Brasil 2950, Casilla 4059, Valpara\'{\i}so, Chile.
}

\begin{abstract}
In this paper we consider a new approach to unify inflation and the late universe with dark energy and dark matter formulated in a model that includes a non-Riemannian metric independent measure and a scalar field with spontaneously broken scale symmetry. Here first of all inflation is possible, which is then followed by a reheating oscillating period and this leads to the formation of all kind of particles, including fermions, which as the universe expands can contribute to the dark energy and the to the dark matter of the universe. 
During the inflationary epoch, we find different constraints on the parameter space associated to the effective potential of the scalar field from the observational data.
After reheating the scalar field retraces its trajectory in field space but now the scalar field potential can be drastically modified by the effect of the fermions. In this sense, the present dark energy with its very small value in comparison to the inflationary phase which  can be adjusted by choosing appropriately the parameter space of couplings  of the Riemannian and non Riemannian measures to the fermions. 
\end{abstract}
\keywords{Spontaneously broken scale symmetry; fermions and dark
energies; domain structure;
accelerating universe.}
\maketitle

\section{Introduction}
The accelerated expansion of the present universe \cite{accel} revived the idea of Einstein of introducing a cosmological constant term, or of
Dark Energy (DE) of the late universe, for reviews of DE, from the theory point of view see \cite{d.e.}. One of the interesting aspects that have been pointed out is the coincidence, at least in orders of magnitude between Dark Energy and Dark Matter (DM) \cite{coinc}, in fact in this paper we will find that DE and  DM may have a common origin.
One approach to DE is to assume that the accelerated expansion is driven by a scalar field with a potential rather than just a cosmological constant \cite{quint}, in our case , a scalar field will play a role, but fermionic matter will also be important.

Of course a much larger cosmological constant is thought to be present in the early universe, driving an even much faster accelerated expansion of the universe, the inflationary phase \cite{inflation}, which solves many puzzles and is though to be also the source of the primordial perturbations. The inflation then ends at some point and this is then followed by particle creation which produces all the matter in the universe, after 13 billion years we have then again a universe dominated by something the cosmological constant mentioned at the beginning which is much smaller than the inflationary cosmological constant.

On the other hand,  sterile neutrinos can be 
 DM and these particles are the  basis for some 
models as also for the generation of 
 DE\cite{Neutr-dark}.
An idea has been
suggested in Refs.\cite{FNW-1,FNW-2}  in which a coupling of the
neutrinos with a  scalar field produces a dark energy density that depends on the neutrino mass.
In this approximation  the neutrino mass  is related to 
 the density of the background nonrelativistic neutrinos in a power-law relation.
We mention that in the model described in Ref.\cite{FNW-1}, the energy density associated to the  neutrinos
is small in relation to  the energy density in the total dark energy
sector.

The generation of dark energy by fermions is studied in the context of the Two Measures Theory, where the traditional measure of integration  $\sqrt{-g}$ is used in some parts of the Lagrangian of the theory, but is replaced by a metric independent measure $\Phi$ in other parts of the Lagrangian  \cite{GK1}, \cite{GK2},\cite{GK3}. then to this idea, the requirement of scale invariance was added im  \cite{G1}, where the spontaneous scale symmetry was was generated by the integration of the degrees of freedom defining the modified measure. Scale invariance modified measures models for Braneworlds were studied in Ref.\cite{G2}.

The subject of unifying the early inflation with the late DE was first studied in the Two Measure Theory in \cite{tfr} by introducing curvature square terms, then this approach was used to obtain emergent universe solutions that evolve from an emergent non singular phase to inflation and then to the present DE phase,
\cite{conSergio1}, and then have shown that similar effect can take place even without curvature square terms \cite{conSergio2} by showing thar the two measures theory can induce K essence terms when considering the Einstein frame, that can also support the emergent solutions and then combined effects of the induced K essence terms and curvature square terms to obtain an Emergent Cosmology, Inflation and Dark Energy were considered in \cite{EMILSVETLANA}. The models  \cite{tfr}, \cite{conSergio1},  \cite{conSergio2} and \cite{EMILSVETLANA}, are non oscillating ones, so to account for reheating after inflation, a curvaton field must be added, this has been done in \cite{curvaton1}, \cite{curvaton2} 
These kind of models provide also the possibility to incorporate inflation, early dark energy and late dark energy, which can be of use in the resolution of the $H_0$ tension \cite{H0}, which was then generalized  \cite{H0DM} to account for the non singular origin of the universe, the $H_0$ problem and the DM as originating from the scalar field due to its K essence behavior. 

In a previous publication, dark energy generation from fermions was considered in the Two Measure Theory\cite{Guendelman:2006ji}, but inflation was not studied, so it was not a full description of the full history of the universe. Furthermore, the model studied here not only  accounts for both inflation and dark energy, but also the effective potential has an oscillating region that allows particle creation and reheating, which allows for the creation of the fermions themselves that will ultimately lead to a modification of the effective potential, which in the same region in scalar field space where inflation took place,  now provides an energy density of the order of $(10^{-3}ev)^4$ corresponding to the DE for the late universe, while the fermions provide the DM as well.

In this article, we want to describe  the unification of the early and present universe. To describe the early universe we assume an inflationary stage dominates by a scalar field from an effective potential without the present of fermions. 
Besides, we want to explore the possibility that the matter produced in the reheating after inflation can have a role in the modification of the cosmological constant of the early universe. 
We explore in particular the possibility that fermions, like neutrinos,
or sterile neutrinos could modify the original inflaton field potential generating a new effective potential, and this same inflaton field   could then drive the late universe expansion. In addition, we can account also for the dark matter of the present universe from these fermions assuming that this matter is also produced during the reheating mechanism.

We organize our paper as follows: In Section \ref{1}, we provide a brief analysis of our model. In this section, we also present the basic equations in the Einstein frame associated with the scalar field and primordial fermions. In Section \ref{2}, we study the inflationary stage of our model. Under the slow-roll approximation, we derive the observational parameters, such as the scalar spectral index, the power spectrum, and the tensor-to-scalar ratio. From these observables, we obtain different constraints on the parameter space of our model. Additionally, in Sections \ref{3} and \ref{4}, we analyze dark energy and dark matter from fermions. Here, we show that it is possible to obtain fermions associated with DE and DM from different energy-momentum tensors. In Section \ref{conc}, we present our conclusions. Finally, we include Appendix \ref{app} for a discussion of the spin connection and the Einstein frame.   We chose units in which  $c=\hbar=1$.

\section{General description}\label{1}

In this section, we will present a brief analysis of the  
 implications of considering both the scalar field and fermion fields. Thus, 
in order to describe  our model, we consider that 
 the general action can be written as \cite{Guendelman:2006ji}
\begin{eqnarray}
&S&=\int d^{4}x e^{\alpha\phi /M_{p}}(\Phi +b\sqrt{-g})
\left[-\frac{1}{\kappa}R(\omega ,e) +
\frac{1}{2}g^{\mu\nu}\phi_{,\mu}\phi_{,\nu}\right]
\nonumber\\
&& -\int d^{4}x e^{2\alpha\phi /M_{p}}[\Phi V_{1} +\sqrt{-g}V_{2}]
-\int
d^{4}x\sqrt{-g}\frac{1}{4}g^{\alpha\beta}g^{\mu\nu}F_{\alpha\mu}F_{\beta\nu}
\nonumber\\
&& +\int d^{4}x e^{\alpha\phi /M_{p}}(\Phi +k\sqrt{-g})
\frac{i}{2}\sum_{i}\overline{\Psi}_{i}
\left(\gamma^{a}e_{a}^{\mu}\overrightarrow{\nabla}^{(i)}_{\mu}-
\overleftarrow{\nabla}^{(i)}_{\mu}\gamma^{a}e_{a}^{\mu}\right)\Psi_{i}
\nonumber\\
     &&-\int d^{4}xe^{\frac{3}{2}\alpha\phi /M_{p}}
\left[(\Phi +h_{E}\sqrt{-g})\mu_{E}\overline{E}E +(\Phi
+h_{n}\sqrt{-g})\mu_{n}\overline{n}n \right], \label{totaction}
\end{eqnarray}
in which  the measure fields $\varphi_{a}$ are utilized for definition of
the measure $\Phi=\varepsilon^{\mu\nu\alpha\beta}\varepsilon_{abcd}\partial_{\mu}\varphi_{a}
\partial_{\nu}\varphi_{b}\partial_{\alpha}\varphi_{c}
\partial_{\beta}\varphi_{d}.$  Therefore, 
 the quantity  $\Psi_{i}$ ($i=n,E$ neutrinos, electrons) is the general notation for the
primordial fermion fields $n$ and $E$; the variable $F_{\alpha\beta}$ is defined as 
$F_{\alpha\beta}=\partial_{\alpha}A_{\beta}-\partial_{\beta}A_{\alpha}$ and the parameters
$\mu_{n}$ and $\mu_{E}$ are the mass parameters. Besides, we have 
$\overrightarrow{\nabla}^{(n)}_{\mu}=\vec{\partial}+
\frac{1}{2}\omega_{\mu}^{cd}\sigma_{cd}$,
$\overrightarrow{\nabla}^{(E)}_{\mu}=\vec{\partial}+
\frac{1}{2}\omega_{\mu}^{cd}\sigma_{cd}+ieA_{\mu}$;
 $R(\omega ,V)
=e^{a\mu}e^{b\nu}R_{\mu\nu ab}(\omega)$ is the scalar curvature;
$e_{a}^{\mu}$ and $\omega_{\mu}^{ab}$ are the vierbein and
spin-connection; $g^{\mu\nu}=e^{\mu}_{a}e^{\nu}_{b}\eta^{ab}$ and
$R_{\mu\nu ab}(\omega)=\partial _{\mu}\omega_{\nu
ab}+\omega^{c}_{\mu a}\omega_{\nu cb}-(\mu \leftrightarrow\nu)$.
The quantities 
$V_{1}$ and $V_2$ are constants with the dimensionality
$(mass)^4$. In addition, the  constants $b, k, h_{n}, h_{E}$  are non specified
dimensionless real parameters. 
In relation to the parameter  $\alpha$, we have that it is a real parameter
which   is   positive and dimensionless. We will determine the relative signs of the kinetic terms and mass terms in the above lagrangian  comparing only the terms that are multiplied by the conventional measure, and since $k$ will be determined to negative, we will choose the corresponding coefficients of the mass terms , like 
 $\mu_{E}$  and  $\mu_{n}$ to be negative. 

We note that the general action given by Eq.(\ref{totaction}) is invariant under the global scale
transformations
\begin{eqnarray}
    &&e_{\mu}^{a}\rightarrow e^{\theta /2}e_{\mu}^{a}, \quad
\omega^{\mu}_{ab}\rightarrow \omega^{\mu}_{ab}, \quad
\varphi_{a}\rightarrow \lambda_{ab}\varphi_{b}\quad
A_{\alpha}\rightarrow A_{\alpha}, \nonumber
\\
 &&\phi\rightarrow
\phi-\frac{M_{p}}{\alpha}\theta ,\quad \Psi_{i}\rightarrow
e^{-\theta /4}\Psi_{i}, \quad \overline{\Psi}_{i}\rightarrow
e^{-\theta /4} \overline{\Psi}_{i},\,\,\,\,\, \det(\lambda_{ab})=e^{2\theta}.
%\nonumber
%\\
% &&\mbox{where} \quad \theta =\mbox{const}, \quad \lambda_{ab}=\mbox{const} %\quad \mbox{and}
% \quad 
\label{stferm}
\end{eqnarray}

In relation to the two measure theory we have that it is a generally coordinate invariant theory, where the action
becomes \cite{GK1}-\cite{GK3}
\begin{equation}
    S = \int L_{1}\Phi d^{4}x +\int L_{2}\sqrt{-g}d^{4}x,
\label{S}
\end{equation}
where there are two Lagrangians $ L_{1}$ and $L_{2}$ and two
measures of integration: the standard $\sqrt{-g}$ and the new one
$\Phi$. Here, we have that  the Lagrangian $L_1$ satisfies the relation \cite{GK1}
\begin{equation}
B^{\mu}_{a}\partial_{\mu}L_{1}=0, \,\,\,\,\mbox{and then}\,\,\,\,L_1=M^4=\mbox{const.} \label{var-phi}
\end{equation}
where $B^{\mu}_{a} = \varepsilon^{\mu\nu\alpha\beta}\varepsilon_{abcd}\partial_{\nu}\varphi_{b}\partial_{\alpha}\varphi_{c}
\partial_{\beta}\varphi_{d}$ and since the determinant of the matrix $B^{\mu}_{a}$ is proportional to the measure, it follows that for a non singular measure we have  that $\partial_{\mu}L_{1}=0$.

From the action given by Eq.(\ref{totaction}), we have a first order formalism that
contain terms proportional to $\partial_{\mu}\chi$, where $\chi=\Phi/\sqrt{-g}$. Thus, we obtain  that the
space-time non-Riemannian and equations of motion - non canonical.

In order to obtain the Einstein frame, we can define 
a new set of variables ($\phi$ and $A_{\mu}$
remain unchanged) such that 
\begin{eqnarray}
&&\tilde{e}_{a\mu}=e^{\frac{1}{2}\alpha\phi/M_{p}}(\chi
+b)^{1/2}e_{a\mu}, \quad
\tilde{g}_{\mu\nu}=e^{\alpha\phi/M_{p}}(\chi +b)g_{\mu\nu},
\nonumber\\
&&\Psi^{\prime}_{i}=e^{-\frac{1}{4}\alpha\phi/M_{p}} \frac{(\chi
+k)^{1/2}}{(\chi +b)^{3/4}}\Psi_{i} , \quad \mbox{with}\,\,\,\,i=n,E .\label{ctferm}
\end{eqnarray}
Note that $\tilde{e}_{a\mu}$,
$\tilde{g}_{\mu\nu}$, $n^{\prime}$ and $E^{\prime}$ are invariant
under the scale transformations (\ref{stferm}).

In the Einstein frame  the  equations  take the standard General Relativity  form given by 
\begin{equation}
G_{\mu\nu}(\tilde{g}_{\alpha\beta})=\frac{\kappa}{2}T_{\mu\nu}^{eff},
 \label{gef}
\end{equation}
in which the tensor  $G_{\mu\nu}(\tilde{g}_{\alpha\beta})$ corresponds to  the Einstein
tensor in the Riemannian space-time with the metric
$\tilde{g}_{\mu\nu}$ and the quantity $\kappa=8\pi/(M_p^2)$ with $M_p$ the Planck mass. Also,  the energy-momentum tensor
$T_{\mu\nu}^{eff}$ is defined as
\begin{eqnarray}
T_{\mu\nu}^{eff}&=&\phi_{,\mu}\phi_{,\nu}-\frac{1}{2}
\tilde{g}_{\mu\nu}\tilde{g}^{\alpha\beta}\phi_{,\alpha}\phi_{,\beta}
+\tilde{g}_{\mu\nu}V_{eff}(\phi;\chi)
%\nonumber\\
%&+&
+T_{\mu\nu}^{(em)}
+T_{\mu\nu}^{(ferm,can)}+T_{\mu\nu}^{(ferm,noncan)},
 \label{Tmn}
\end{eqnarray}
where the effective potential associated to scalar field and $\chi$ is given by
\begin{equation}
V_{eff}(\phi,\chi)=
\frac{b\left(M^{4}e^{-2\alpha\phi/M_{p}}+V_{1}\right)
-V_{2}}{(\chi +b)^{2}}. \label{Veff1}
\end{equation}
Besides, the quantity $T_{\mu\nu}^{(em)}$ corresponds to the canonical energy momentum tensor of the
electromagnetic field and the tensor $T_{\mu\nu}^{(ferm,can)}$ represents the canonical
energy momentum tensor for primordial fermions $n^{\prime}$ and
$E^{\prime}$ in curved space-time. The quantity $T_{\mu\nu}^{(ferm,noncan)}$ denotes the
 noncanonical contribution of the fermions into the energy
momentum tensor and it is defined as\cite{Guendelman:2006ji}
\begin{equation}
 T_{\mu\nu}^{(ferm,noncan)}=-\tilde{g}_{\mu\nu}\Lambda_{dyn}^{(ferm)},
 \label{Tmn-noncan}
\end{equation}
where $\Lambda_{dyn}^{(ferm)}$ is given by
\begin{equation}
\Lambda_{dyn}^{(ferm)}\equiv Z_{n}(\chi)m_{n}(\chi)
\overline{n^{\prime}}n^{\prime}+
Z_{E}(\chi)m_{E}(\chi)\overline{E^{\prime}}E^{\prime},
\label{Lambda-ferm}
\end{equation}
and $Z_{i}(\chi)$ and $m_{i}(\chi)$ ($i=n^{\prime},E^{\prime}$)
are defined as
\begin{equation}
Z_{i}(\chi)\equiv \frac{(\chi -\chi^{(i)}_{1})(\chi
-\chi^{(i)}_{2})}{2(\chi +k)(\chi +h_{i})}, \qquad
m_{i}(\chi)= \frac{\mu_{i}(\chi +h_{i})}{(\chi +k)(\chi
+b)^{1/2}}, \label{Zeta&m}
\end{equation}
where 
\begin{equation}
\chi_{1,2}^{(i)}=\frac{1}{2}\left[k-3h_{i}\pm\sqrt{(k-3h_{i})^{2}+
8b(k-h_{i}) -4kh_{i}}\,\right].
 \label{zeta12}
\end{equation}

It is important  to mention that the noncanonical contribution given by $T_{\mu\nu}^{(ferm,noncan)}$ of the
fermions into the energy momentum tensor produces  a cosmological constant like term but it is proportional
to fermion densities
$\overline{\Psi}^{\prime}_{i}\Psi^{\prime}_{i}$ \,
($i=n^{\prime},E^{\prime}$), and  this is why we will refer to it as
"dynamical fermionic $\Lambda$ term".

The  scalar field $\chi$ given by $\chi=\Phi/\sqrt{-g}$,
is determined in terms  of the scalar field $\phi$ and
$\overline{\Psi}^{\prime}_{i}\Psi^{\prime}_{i}$ \,
($i=n^{\prime},E^{\prime}$) by the following equation\cite{Guendelman:2006ji}
\begin{equation}
\frac{1}{(\chi
+b)^{2}}\left\{(b-\chi)\left[M^{4}e^{-2\alpha\phi/M_{p}}+
V_{1}\right]-2V_{2}\right\}=
\Lambda_{dyn}^{(ferm)}. \label{constraint}
\end{equation}
In the following, we will study the inflationary stage in which we will consider the absence of fermions during this epoch.

\section{Inflation}\label{2}

In this section we will analyze the inflationary era where we do not consider fermions so that $\Lambda_{dyn}^{(ferm)}=0$. Thus, from Eq.(\ref{constraint})  we find that the quantity $\chi$ in terms of the scalar field becomes 
\begin{equation}
\chi =\chi_{0}(\phi)\equiv b-\frac{2V_{2}}
{V_{1}+M^{4}e^{-2\alpha\phi/M_{p}}}. \label{zeta-without-ferm}
 \end{equation}
Now replacing this equation into Eq.(\ref{Veff1}) we obtain that 
the effective potential of
the scalar field $\phi$ in absence of fermions results
\begin{equation}
V_{eff}(\phi)\equiv
V_{eff}(\phi;\chi_{0})|_{\overline{\psi^{\prime}}\psi^{\prime}=0}
=\frac{[V_{1}+M^{4}e^{-2\alpha\phi/M_{p}}]^{2}}
{4[b\left(V_{1}+M^{4}e^{-2\alpha\phi/M_{p}}\right)-V_{2}]}.
\label{Veffvac}
\end{equation}

\begin{figure*}[!hbtp]
     \centering	\includegraphics[width=0.4
\textwidth,keepaspectratio]{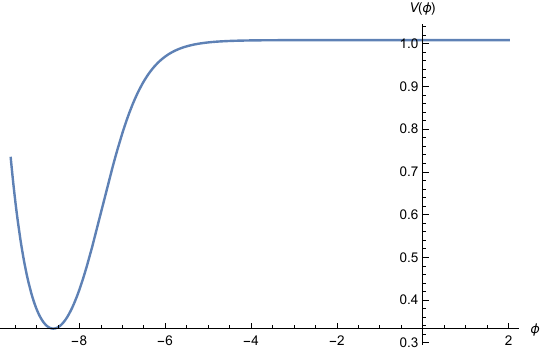}
\caption{A schematic representation of the effective potential $V(\phi)$ as a function of the scalar field $\phi$, given by Eq. (\ref{Veffvac}). The flat region corresponds to the inflationary phase described by Eq. (\ref{Vfl}). At the minimum, the mechanism of reheating, and also  the production of fermions occurs.   }
\label{Fig1}
\end{figure*}

In order to analyze the inflationary period, we can consider that  
 the scalar field start from very high positive values ($\phi\rightarrow+\infty$), with a flat region given by 
\begin{equation}
V_{eff}
(\phi)\simeq\frac{V_1^2}{4(bV_1-V_2)},\,\,\,\,\mbox{with}\,\,\,\,\,bV_1>V_2.\label{Vfl}
\end{equation}

 Fig.\ref{Fig1} shows the effective potential as a function of the scalar field defined by Eq.(\ref{Veffvac}). For this potential, we note that the flat region refers  to the inflationary phase defined by Eq.(\ref{Vfl}). Furthermore, at the minimum of this effective potential, the mechanism of reheating occurs. This process involves the perturbative decay of the oscillating field $\phi$ at the end of the inflationary scenario, which generates the reheating of the universe\cite{reh}.  In this context, various particles, including fermions, are produced at the minimum of the effective potential.  
Here we have used the values; $\alpha=1$, $V_1=(1.1/3) M_p^4$, $V_2=(1/3)M_p^4$, $M=0.01M_p$ and $b=1$, respectively.

Now, expanding the  effective potential (\ref{Veffvac}), we obtain that the next term is given by   
\begin{equation}
V_{eff}(\phi)\simeq\frac{V_1^2}{4(bV_1-V_2)}+\frac{V_1M^4}{2(bV_1-V_2)}\left[1-\frac{bV_1}{2(bV_1-V_2)}\right]e^{-2\alpha\phi/M_p}+....,\label{Poap}
\end{equation}
with $bV_1/2<V_2$, in order to obtain a decrease of the effective potential as we move to smaller values of the scalar field during the inflationary epoch. However, we also have that $bV_1>V_2$, then the condition on the parameter $V_2$ becomes $bV_1/2<V_2<bV_1$.

This choice $bV_1/2<V_2$ differs from that of Ref.\cite{Guendelman:2006ji}, where inflation was not considered and the evolution was in the other direction.
In contrast, here, the high values of the scalar field will be used twice, once for inflation and then, after  reheating, the same scalar field goes back in the other direction, but now feeling a modified potential modified by the particles created during the reheating.
%For the early dark energy we can assume that the scalar field %$\phi\rightarrow-\infty$, and then  the effective potential is reduced to
%\begin{equation}
%V_{eff}^{(0)}(\phi)\simeq\frac{M^4}{4b}\,e^{-2\alpha\phi/M_{p}}.
%\end{equation}

In the context of the slow roll approximation and considering a flat Friedmann-Roberson-Walker metric,  we have that the  equations of motion are reduced to
\begin{equation}
3H^2=\kappa\rho\simeq\kappa \,V_{eff},\,\,\,\,\mbox{and}\,\,\,\,\,3H\dot{\phi}\simeq -\frac{\partial V_{eff}}{\partial \phi},
\end{equation}
where $H=\dot{a}/a$ denotes the Hubble parameter and the quantity $a$ the scale factor. Here, in the absence of fermions, we have considered Eq.(\ref{gef}) together with the conservation of the energy-momentum tensor, defined by Eq. (\ref{Tmn}).

Thus, under the slow roll approximation from Eq.(\ref{Poap}), we find that the scalar field in terms of the cosmological time can be written as
\begin{equation}
e^{\alpha\phi/M_p}\sqrt{(V_{01}+V_{02}e^{-2\alpha\phi/M_p})}\left(e^{\alpha\phi/M_p}+\frac{\sqrt{V_{02}}\mbox{ArcSinh}[e^{\alpha\phi/M_p}\sqrt{V_{01}/V_{02}}]}{\sqrt{V_{01}}\sqrt{1+e^{2\alpha\phi/M_p}V_{01}/V_{02}}}\right)=\frac{4V_{02}(\alpha/M_p)^2}{\sqrt{3\kappa}}\,t+C_0,\label{ap1}
\end{equation}
where $C_0$ corresponds to an integration constant. Here the constants $V_{01}$ and $V_{02}$ are defined as
\begin{equation}
V_{01}=\frac{V_1^2}{4(bV_1-V_2)},\,\,\,\,\mbox{and} \,\,\,\,\,\,V_{02}=\frac{V_1M^4}{2(bV_1-V_2)}\left[1-\frac{bV_1}{2(bV_1-V_2)}\right]<0,
\label{VV1}\end{equation}
respectively. From Eq.(\ref{ap1}), we note that for $\phi\rightarrow \infty$, the solution for the scalar field in terms of time is approximated  to
$$
e^{2\alpha\phi/M_{p}}\sim \frac{4V_{02}(\alpha/M_p)^2}{\sqrt{3\kappa V_{01}}} t + C'_0,
$$
where $C_0'=C_0/\sqrt{V_{01}}$.

By introducing the slow roll parameters $\epsilon$ and $\eta$, we have

\begin{equation}
\epsilon=\frac{1}{2\kappa} \left[\frac{\partial V_{eff}/\partial\phi}{V_{eff}}\right]^2,\,\,\,\,\,\mbox{and}\,\,\,\,\eta=\frac{1}{\kappa}\left[\frac{\partial^2V_{eff}/\partial\phi^2}{V_{eff}}\right].
\end{equation}
Thus,  considering that the end of inflation taken place when $\epsilon(\phi=\phi_e)=1$, we find the  value at the end of inflation becomes
\begin{equation}
e^{-2\alpha\phi_e/M_p}=-\left[\frac{V_{01}}{V_{02}}\right]\,\frac{1}{[\frac{\sqrt{2}\alpha}{\sqrt{\kappa}M_p}+1]}.\label{ff}
\end{equation}
Here $\phi_e$ denotes the value of the scalar field at the end of inflationary epoch and recalled that $V_{02}<0$.

Also, the number of e-folds $N$ between two different values of the times $t$ and $t_e$ results
\begin{equation}
N=\int_t^{t_e}H\,dt\simeq\kappa\int_{\phi_e}^\phi\,\frac{V_{eff}}{(\partial V_{eff}/\partial\phi)}\,d\phi,
\end{equation}
where $t_e$ denotes the end of the inflationary epoch. Thus, we find that the number of $e-$folds as a function of the scalar field can be written as 
\begin{equation}
N=\left(\frac{\kappa M_p}{2\alpha |V_{02}|}\right)\left(\left[\frac{M_pV_{01}e^{2\alpha\phi/M_{p}}}{2\alpha}+V_{02}\phi\right]-\left[\frac{M_pV_{01}e^{2\alpha\phi_e/M_{p}}}{2\alpha}+V_{02}\phi_e\right]\right).
\end{equation}
 In this context, we can determine the scalar field as a function of the number of $e-$ folds as
 \begin{equation}
\phi(N)=\frac{\tilde{N}}{V_{02}}-
\frac{M_p \mbox{ProducLog}[(V_{01}/V_{02})e^{2\alpha\tilde{N}/(M_p V_{02})}]}{2\alpha},
 \end{equation}
in which the quantity $\tilde{N}$ is defined as
$$
\tilde{N}=\left(\frac{2\alpha |V_{02}|}{\kappa M_p}\right)\,N+\left[\frac{M_pV_{01}e^{2\alpha\phi_e/M_{p}}}{2\alpha}+V_{02}\phi_e\right],
$$
where the scalar field at the end of inflation $\phi_e$ is given by Eq.(\ref{ff}). 
Besides, the ProductLog function corresponds to  the product logarithm, also called the Omega function or  Lambert W function. This function is a multivalued function, and it satisfy  the converse relation of the function $f(w) = we^w$, see Ref.\cite{Prod}

On the other hand, the power spectrum $\mathcal{P}_R$ of the curvature perturbations, in the context of the slow roll approximation becomes\cite{p1}
\begin{equation}
\mathcal{P}_R=\left(\frac{H^2}{\dot{\phi}^2}\right)\,\left(\frac{H}{2\pi}\right)^2\simeq\frac{\kappa^3}{12\pi^2}\,\left[\frac{V_{eff}^3}{(\partial V_{eff}/\partial\phi)^2}\right],\label{Pr}
\end{equation}
and the scalar spectral index $n_s$ results \cite{p1}
\begin{equation}
n_s-1=-6\epsilon+2\eta,\label{ns}
\end{equation}
and the tensor to scalar ratio $r$ is defined as \cite{p1}
\begin{equation}
r=\frac{8}{\kappa}\,\left[\frac{\partial V_{eff}/\partial\phi}{V_{eff}}\right]^2.\label{r}
\end{equation}

From Eq.(\ref{Pr}) we find that the scalar power spectrum results
\begin{equation}
\mathcal{P}_R\simeq k_1\,e^{4\alpha\phi/M_p}\,\left[V_{01}+V_{02}e^{-2\alpha\phi/M_p}\right]^3,
\end{equation}
where the constant $k_1$ is defined as
$$
k_1=\frac{\kappa^3M_p^2}{48\pi^2\alpha^2V_{02}^2}.
$$
Also, we determine that the scalar spectral index in terms of the scalar field from Eq.(\ref{ns}) results 
\begin{equation}
n_s-1=-\left[\frac{4\alpha^2 |V_{02}|}{\kappa M_p^2}\right](|V_{02}|+2V_{01}e^{2\alpha\phi/M_p})\,(V_{01}+V_{02}e^{2\alpha\phi/M_p})^{-2}.
\end{equation}
From Eq.(\ref{r}) we find that the tensor to scalar ratio $r$ as a function of the scalar field becomes
\begin{equation}
r=r_0\,e^{-4\alpha\phi/M_p}\,(V_{01}+V_{02}e^{2\alpha\phi/M_p})^{-2},
\end{equation}
where the constant $r_0$ is defined as
$$
r_0=\frac{32\alpha^2V_{02}^2}{\kappa M_p^2}.
$$
In particular assuming the value of the parameter $\alpha=1$ together with the cosmological parameters $\mathcal{P}_R=2.2\times 10^{-9}$ and $n_s=0.968$ at  $N=60$ (see \cite{Planck:2018vyg}), then numerically we obtain that the parameters $V_{01}$ and $V_{02}$ result
\begin{equation}
V_{01}\simeq4,034\times 10^{-4}M_p^4,\,\,\,\,\,\,\mbox{and}\,\,\,\,\,
|V_{02}|=5,286\times10^{-5}M_p^4,\label{cons}
\end{equation}
respectively. In addition, using these values,  we find that from these values  the tensor to scalar ratio $r$ at $N=60$, becomes $r\simeq0.01299$.

In relation to the original parameters, we note that by substituting the constraints given by Eq.(\ref{cons}) into Eq. (\ref{VV1}), we find that the parameter 
$M$, obtained from the condition (\ref{var-phi}), can be expressed in terms of the parameters 
$V_1$ and $b$ as follows:
\begin{equation}
M\simeq\left(\frac{6.3\times 10^{-3}V_1}{\frac{b}{125V_1}-1}\right)^{1/4},\label{M}
\end{equation}
with the condition  $b>125V_1$ to ensure  a real solution for the parameter $M$. Besides, we note that for the special situation in which the ratio $b/125V_1\gg 1$, the parameter   $M$ becomes  $M\simeq V_1^{1/2}\,b^{-1/4}$. Additionally, we  find that the equation for $V_2$ in terms of $V_1$ and $b$ is given by
\begin{equation}
V_2\simeq bV_1-625V_1^2. 
\end{equation}
Here we need to consider that $b>625V_1$ to obtain a positive value of the parameter $V_2$. Thus, this condition on the parameter $b$ is more strong that the condition $b>125V_1$ obtained from Eq.(\ref{M}) to find a real solution of the parameter $M$.
As before, here we have considered Eqs.(\ref{VV1}) and (\ref{cons}), respectively. In addition, we note that  for the case in which the parameter  $V_1\ll1$, we have that $V_2\sim b\,V_1$.

After of the inflationary stage, the scalar field oscillates above the minimum of the effective potential and during this process occurs the particles production and in particular the fermions. In particular, we will look at the effect of the fermions as they get diluted and how they affect the dark energy in the late universe. 

In the following, we will analyze the effect of the fermions as  dark energy  and dark matter.

\section{dark energy from fermions}\label{3}

In this section we will study the accelerated expansion of the present universe produced from the effective potential associated to the  fermions generated during the inflationary stage.

For dust, the neutrino has zero momenta, and then we have
$\overline{n}^{\prime}n^{\prime}= u^{\dagger}u$ where $u$ corresponds to the
large component of the Dirac spinor $n^{\prime}$. From the space
components of the 4-current
$\tilde{e}_{a}^{\mu}\overline{n}^{\prime}\gamma^{a}n^{\prime}$
equal zero and assuming  the 4-current conservation  we have 
$\overline{n}^{\prime}n^{\prime}= u^{\dagger}u
=\frac{const}{a^{3}}$ where $a=a(t)$ as before is the scale factor.
Thus, from Eq.
(\ref{constraint}) we have
\begin{equation}
\frac{(b-\chi )\left[M^{4}e^{-2\alpha\phi/M_{p}}+
V_{1}\right]-2V_{2}}{(\chi +b)^{3/2}}=\frac{(\chi
-\chi^{(N)}_{1})(\chi -\chi^{(N)}_{2})}{(\chi +k)^{2}}\mu_{N}
\frac{const}{a^{3}}. \label{constraint-toy}
\end{equation}

We note that
there is a solution where the decaying fermion
contribution $u^{\dagger}u\sim \frac{const}{a^{3}}$ is not trivial to the  right hand side of Eq.(\ref{constraint-toy}). In this sense, 
 to generate the dark energy from the fermions, we can consider the asymptotic behavior $\chi\rightarrow -k$ in such a way
that $(\chi +k)^{-2} \propto a^{3}$, then the r.h.s. of
(\ref{constraint-toy}) became a constant. We note that the described regime denotes a very
unexpected state of the primordial fermions.

To obtain the total effective potential $U_{eff}(\phi)$ in terms of the scalar field associated with fermions during the current universe, we have the following  
\begin{equation}
U_{eff}=V_{eff}- \Lambda_{dyn}^{(ferm)}\simeq\frac{V_{2}-kV_{1}}{(b-k)^{2}}-\frac{k}{(b-k)^{2}}M^{4}e^{-2\alpha\phi/M_{p}}+{\cal
O}\left(\chi +k\right).
\end{equation}
Here we have utilized Eqs.(\ref{Veff1}) and (\ref{constraint}) together with  
$\chi\rightarrow -k$.
Notice that this vacuum energy can be very small if  $-k$ is very big.

We note from this total effective potential $U_{eff}$  we have a modified  flat region for positive large $\phi$ associated to dark energy at the present given by 
\begin{equation}
U_{eff}(\phi\rightarrow \infty)=U_{DE}\simeq\frac{V_{2}-kV_{1}}{(b-k)^{2}}.\label{Ue}
\end{equation}
In addition, we can define the density parameter associated  to the dark energy $\Omega_{DE}=\kappa\rho_{DE}/3H^2\simeq\kappa U_{eff}/3H^2$. Thus, evaluating the parameter $\Omega_{DE}$ at the present time i.e., $\Omega_{{DE}_0}$, we find
\begin{equation}
\Omega_{{DE}_0}\simeq\frac{\kappa}{3H_0^2}\left(\frac{V_{2}-kV_{1}}{(b-k)^{2}}\right).
\end{equation}
By considering that the Hubble parameter at the present time, $H_0\approx10^{-61}M_p$, we obtain the constraint
\begin{equation}
\frac{V_{2}-kV_{1}}{(b-k)^{2}}\approx 2\times 10^{-122}\frac{M_p^2}{\kappa},\label{36}
\end{equation}
where we have used that at the present the parameter $\Omega_{{DE}_0}\simeq0.7$.

Besides, for very large values of $k$ and considering Eq.(\ref{Ue}), we can obtain an approximation for the parameter $k$ and it becomes  of  order 
\begin{equation}
|k|\sim \frac{V_1}{U_{DE}}\sim 10^{119}.\label{kk}
\end{equation}
Here, we have assumed  that $V_1\sim V_2$, $b\sim \mathcal{O}(1)$, and,  from Eq.(\ref{cons}) $V_{01}\sim V_1 \sim 10^{-4}M_p^4$, obtained during the inflationary epoch, together with  the value of the dark energy at the present epoch,   $U_{DE}\sim 10^{-123}M_p^4$.

\section{dark matter from fermions}\label{4}

In this section, we will analyze the dark matter associated with the high-density fermions (localized) in our present universe generated during the inflationary epoch. 

From the constraint equation given by Eq.(\ref{constraint}), the right hand size is proportional to the fermions densities. In this context, $\Lambda_{dyn}^{(ferm)}$ given by Eq.(\ref{Lambda-ferm})  becomes very large unless the coefficient $Z_{i}$ defined by Eq.(\ref{Zeta&m}) goes to zero. Thus, to obtain the coefficient $Z_i\rightarrow 0$, we need to consider that $\chi\rightarrow \chi_{1,2}^{(i)}$ where $\chi_{1,2}^{(i)}$ are defined by Eq.(\ref{zeta12}).

By considering that
in the studied regime masses (\ref{Zeta&m}) of the
non-relativistic primordial fermions are very large, we can
neglect their kinetic energy. Thus, after averaging on the microscopic
 contribution to the energy-momentum tensor
$T_{\mu\nu}^{(n,can)}$ we can written \cite{Guendelman:2006ji}
\begin{equation}
<T_{\mu\nu}^{(n',can)}>_{cosm.av.}\simeq\delta_{\mu}^{0}\delta_{\nu}^{0}
\frac{h_{n}-k}{(b-k)^{{1/2}}}\mu_{n}\frac{c}{(\chi +k)a^{3}}, \label{Tmn-can-aver}
\end{equation}
where $c$ is a constant.  In order to obtain that the coefficient $Z_i\rightarrow 0$ and considering only the neutrinos  ( $i=n'$), in which    $\chi\rightarrow \chi_{1,2}^{n'}$ and then we have that  the canonical
energy momentum tensor for fermions $n^{\prime}$ results
\begin{equation}
<T_{\mu\nu}^{(n',can)}>_{cosm.av.}\simeq\delta_{\mu}^{0}\delta_{\nu}^{0}
\frac{h_{n}-k}{(b-k)^{{1/2}}}\mu_{n}\frac{c_{1,2}}{(\chi_{1,2}^{n'} +k)a^{3}}\propto\frac{1}{a^3}. \label{Tmn-can-aver2}
\end{equation}
Here we note that there are two cases for the canonical energy momentum associated to the coefficients $\chi_{1,2}^{n'}$ (solutions) defined by Eq.(\ref{zeta12}). 

Thus, for the first solution  $\chi_{1}^{n'}$ from Eq.(\ref{Tmn-can-aver2}) we find that $<T_{\mu\nu}^{(n',can)}>_{cosm.av.}$ becomes

\begin{equation}
<T_{\mu\nu}^{(n',can)}>_{cosm.av.}\simeq\delta_{\mu}^{0}\delta_{\nu}^{0}
\frac{h_{n}-k}{(b-k)^{{1/2}}}\mu_{n}\frac{c_{1}}{(\chi_{1}^{n'} +k)a^{3}}\propto\frac{1}{a^3}.
\end{equation}
 In the case when consider the second solution for the coefficient  $\chi_{2}^{n'}$  we have that the canonical
energy momentum tensor for fermions is given by
\begin{equation}
<T_{\mu\nu}^{(n',can)}>_{cosm.av.}\simeq\delta_{\mu}^{0}\delta_{\nu}^{0}
\frac{h_{n}-k}{(b-k)^{{1/2}}}\mu_{n}\frac{c_{2}}{(\chi_{2}^{n'} +k)a^{3}}\propto\frac{1}{a^3}.
\end{equation}

 In this context, we consider the energy density related to dark matter to be defined as $\rho_{DM}=<T_{00}^{(n',can)(1)}>_{cosm.av.}+<T_{00}^{(n',can)(2)}>_{cosm.av.}$.
In this sense, we find that the energy density associated to dark matter for large-$k$ can be written as

\begin{equation}
\rho_{DM}\simeq
\frac{h_{n}-k}{(b-k)^{{1/2}}}\mu_{n}\left[\frac{c_{1}}{(\chi_{1}^{n'} +k)}+\frac{c_{2}}{(\chi_{2}^{n'} +k)}\right]\,\frac{1}{a^{3}}\propto\frac{1}{a^3}.
\end{equation}

Defining the density parameter related to the dark matter $\Omega_{DM}=\kappa\rho_{DM}/3H^2$, then we find that at the present time the density parameter becomes
\begin{equation}
\Omega_{{DM}_0}=\kappa\frac{h_{n}-k}{3H_0^2(b-k)^{{1/2}}}\mu_{n}\left[\frac{c_{1}}{(\chi_{1}^{n'} +k)}+\frac{c_{2}}{(\chi_{2}^{n'} +k)}\right],
\end{equation}
where we have used that scale factor at present $a_0=1$. Now considering the density parameter $\Omega_{{DM}_0}\simeq 0.3$ and the Hubble parameter in the present $H_0\approx10^{-61}M_p$, we have
\begin{equation}
\frac{h_{n}-k}{(b-k)^{{1/2}}}\mu_{n}\left[\frac{c_{1}}{(\chi_{1}^{n'} +k)}+\frac{c_{2}}{(\chi_{2}^{n'} +k)}\right]=\frac{3H_0^2\Omega_{{DM}_0}}{\kappa}\approx\left(\frac{10^{-122}M_p^2}{\kappa}\right),\label{const1}
\end{equation}
here we have utilized the value of $k$ given by Eq.(\ref{kk}).

By assuming that the parameter $k$ is very large and the mass parameter $\mu_n$ negative, we find that the constraint given by Eq.(\ref{const1}) is reduced to  
\begin{equation}
|\mu_{n}|\left[\frac{c_{1}}{2}+c_2\right]\approx\left(\frac{10^{-62}M_p^2}{\kappa}\right).\label{const12}
\end{equation}
Here we have used that for large-$k$, the first solution of Eq.(\ref{zeta12}) corresponds to $\chi_{1}^{n'}\rightarrow k$
and the second solution $\chi_{2}^{n'}\rightarrow 0$. In addition, we have considered the valor of $k$ found in  Eq.(\ref{kk}).

\section{Discussion on Possible further Developments}\label{conc}
In this paper we have discussed a unified picture of inflation and the late DE and DM period of the universe based on a scalar field together with  the creation of fermions that in a model with a dilaton field and spontaneously broken scale invariance in the context of  a Two Measures Theory. 

During the inflationary stage  the scalar field produces the accelerated expansion of the universe before the production of exotic fermions. In inflation we have found different constraints related to the effective potential of the scalar field in absence of fermions from the observational parameters. 

In the case of the late universe, the fermions, when diluted can have exotic behavior and can substantially modified the scalar field potential, reducing the very large inflation to the very low DE in the late universe. In order to describe the late universe, we have considered that after reheating the scalar field retraces its trajectory in field space but now the scalar field potential can be drastically modified by the effect of the fermions. From the noncanonical contribution of the fermions associated to the energy momentum tensor we have obtained a part of  the dark energy responsible of the present accelerated universe.  The other contribution to  the dark energy is related to the scalar field contribution which  is also modified by the fermions because of the  solution of the  constraint equation given by Eq.(\ref{constraint-toy}). We have considered that the present dark energy with its very small value in comparison to the inflationary phase which can be  adjusted by choosing appropriately the parameter space of couplings  of the Riemannian and non Riemannian measures to the fermions.

To generate the dark matter we have considered the canonical contribution of fermions related to energy momentum tensor. In these clumped regions, we have that 
 the  matter is more dense. For the dark matter we have assumed two solutions associated to the constraint $\chi\rightarrow \chi_{1,2}^{n'}$ that contributes to the energy density of the dark matter $\rho_{DM}$. In the limit in which  the parameter $k$ is very large, we have obtained a constraint on the mass parameter and the coefficients $c_1$ and $c_2$ from the present density parameter $\Omega_{{DM}_0}$, see Eq.(\ref{const12}).

Of course in the future, there are many issues to be studied, for example, what are these fermions?, 
may be sterile neutrinos?. How this theory works for structure formation?. Since the final value of the DE takes some time as the universe needs to expand so that the asymptotic value of the DE sets in, does this evolution of the DE helps to explain the $H_0$ tension through some kind of Early DE scenario?.

\section*{Acknowledgements}\label{Ack}
EG acknowledges the Pontificia Universidad Cat\'{o}lica de Valpara\'{\i}so for hospitality. He also acknowledges COST action CA23130 - Bridging high and low energies in search of quantum gravity (BridgeQG) and 
CosmoVerse • COST Action CA21136
Addressing observational tensions in cosmology with systematics and fundamental physics for finantial support.
\newpage
\appendix

\section{Connection in the original and Einstein frames}\label{app}
In this appendix,  
we show the dependence of the spin connection
$\omega_{\mu}^{ab}$ on $e^{a}_{\mu}$, $\chi$, $\Psi$ and
$\overline{\Psi}$. Thus, varying the  Eq.(\ref{totaction}) with
respect to $\omega_{\mu}^{ab}$ and considering 
\begin{equation}
R(V,\omega)\equiv
-\frac{1}{4\sqrt{-g}}\varepsilon^{\mu\nu\alpha\beta}\varepsilon_{abcd}
e^{c}_{\alpha}e^{d}_{\beta}R_{\mu\nu}^{ab}(\omega) \label{A1}
 \end{equation}
we find
\begin{eqnarray}
\varepsilon^{\mu\nu\alpha\beta}\varepsilon_{abcd}\left[(\chi
+b) e^{c}_{\alpha}D_{\nu}e^{d}_{\beta}
+\frac{1}{2}e^{c}_{\alpha}e^{d}_{\beta}\left(\chi_{,\nu}+\frac{\alpha}{M_{p}}\phi_{,\nu}\right)\right]+
%\nonumber\\
\frac{\kappa}{4}\sqrt{-g}(\chi
+k)e^{c\mu}\varepsilon_{abcd}\overline{\Psi}
\gamma^{5}\gamma^{d}\Psi=0, \label{A2}
 \end{eqnarray}
in which
\begin{equation}
D_{\nu}e_{a\beta}\equiv\partial_{\nu}e_{a\beta} +\omega_{\nu
a}^{d}e_{d\beta}. \label{A3}
 \end{equation}
The solution of Eq. (\ref{A2}) can be written as
\begin{equation}
\omega_{\mu}^{ab}=\omega_{\mu}^{ab}(e)   +
K_{\mu}^{ab}(e,\overline{\Psi},\Psi) + K_{\mu}^{ab}(\chi, \phi),
\label{A4}
 \end{equation}
where
\begin{equation}
\omega_{\mu}^{ab}(e)=e_{\alpha}^{a}e^{b\nu}\{
^{\alpha}_{\mu\nu}\}- e^{b\nu}\partial_{\mu}e_{\nu}^{a} ,\label{A5}
 \end{equation}
and it corresponds to  the Riemannian part of the spin-connection. Besides, the quantity 
\begin{equation}
K_{\mu}^{ab}(e,\overline{\Psi},\Psi)= \frac{\kappa}{8}\frac{\chi
+k}{\chi +b}
\eta_{cn}e_{d\mu}\varepsilon^{abcd}\overline{\Psi}
\gamma^{5}\gamma^{n}\Psi, \label{A7}
 \end{equation}
denoted the fermionic contribution that differs from the standard expression
\cite{Gasperini} by the factor $\frac{\chi +k}{\chi +b_{g}}$.
Also,  the parameter $K_{\mu}^{ab}(\chi, \phi)$ is defined as
\begin{equation}
K_{\mu}^{ab}(\chi, \phi)=\frac{1}{2(\chi
+b)}\left(\chi_{,\alpha}+\frac{\alpha}{M_{p}}\phi_{,\alpha}\right)(e_{\mu}^{a}e^{b\alpha}-
e_{\mu}^{b}e^{a\alpha}), \label{A6}
 \end{equation}
and it corresponds to  the non-Riemannian part of the spin-connection obtained by
specific features of TMT.

In the Einstein frame,  we determine  that the  variables defined by
Eq.(\ref{ctferm}), the spin-connection read
\begin{equation}
\omega^{\prime ab}_{\mu}=\omega^{ab}_{\mu}(\tilde{e}) +
\frac{\kappa}{8}
\eta_{cn}\tilde{e}_{d\mu}\varepsilon^{abcd}\overline{\Psi}^{\prime}
\gamma^{5}\gamma^{n}\Psi^{\prime}, \label{A7}
 \end{equation}
 which is equivalent to the spin-connection of the Einstein-Cartan
 space-time\cite{Gasperini} with the vierbein $\tilde{e}^{a}_{\mu}$. In the solution of this paper,  the last term given by Eq.(\ref{A7}) is zero.


\begin{thebibliography}{0}

\bibitem{accel} A. G. Riess et al., {\it Astron. J.} {\bf 116}, 1009 (1998);
S. Perlmutter {\it et al.},  {\it Astrophys. J.} {\bf 517}, 565
(1999);
 N. Bahcall, J.P. Ostriker, S.J. Perlmutter and
P.J. Steinhardt, {\it Science} {\bf 284}, 1481 (1999); D.N.
Spergel, {\it et al.} [WMAP collaboration], {\it Astrophys. J.
Suppl. 148}, {\bf 175} (2003); A. C. S. Readhead, {\it et al.}
{\it Astrophys. J.} {\bf 609}, 498 (2004);  J. H. Goldstein et
al., {\it Astrophys.J.} 599,  773-785 (2003); R. Rebolo, {\it et
al.}, astro-ph/0402466;  M. Tegmark, {\it et al.} {\it Phys.Rev.}
{\bf D69}, 103501 (2004); E. Hawkins {\it et al.,} {\it
Mon.Not.Roy.Astron.Soc.} {\bf 346}, 78 (2003);
 W. L. Freedman, {\it et al., Astrophys.J.}
{\bf 553},  47 (2001); R. Daly, S.G. Djorgovski, {\it
Astrophys.J.} {\bf 597}, 9, (2003).


\bibitem{d.e.} S.M.Carroll, {\it Living Rev. Rel.} {\bf 4}, 1 (2001);
 A. Vilenkin, hep-th/0106083; P.J.E. Peebles and  B. Ratra,
 Rev. Mod. Phys. {\bf 75}, 559 (2003);
T. Padmanabhan, Phys. Rep. {\bf 380}, 235 (2003); V. Sahni, {\it
AIP Conf.Proc.} {\bf 782}, 166 (2005).

\bibitem{coinc}
I. Zlatev, L Wang and P. Steinhardt, {\it Phys. Rev. Lett.} {\bf
82}, 896 (1999).

\bibitem{quint} C. Wetterich, {\it Nucl. Phys.}  {\bf B302}, 668
(1988); B. Ratra and P.J.E. Peebles, {\it Phys. Rev.} {\bf D37},
3406 (1988); P.J.E. Peebles and B. Ratra, {\it Astrophys. J.} {\bf
325}, L17 (1988); R. Caldwell, R. Dave and P. Steinhardt, {\it
Phys. Rev. Lett.} {\bf 80}, 1582 (1998); N. Weiss, {\it Phys.
Lett.} {\bf B197}, 42 (1987); Y. Fujii and T. Nishioka, {\it Phys.
Rev.} {\bf D42}, 361 (1990); M.S. Turner and M. White, {\it Phys.
Rev.} {\bf D56}, R4439 (1997); P. Ferreira and M.Joyce, {\it Phys.
Rev. Lett.} {\bf 79}, 4740 (1997); {\it Phys. Rev.} {\bf D58},
023503 (1998). E. Copeland, A. Liddle and D. Wands, {\it Phys.
Rev.} {D57}, 4686 (1998);
 P. Steinhardt, L Wang and I. Zlatev, {\it Phys. Rev.} {\bf
D59}, 123504 (1999).

%\bibitem{amend}
%L. Amendola, {\it Phys. Rev.} {\bf D62}, 043511 (2000).

%\bibitem{vamp}
%J.A.Casas, J. Garcia-Bellido and M. Quiros, {\it Class. Quant.
%Grav.} {\bf 9}, 1371 (1992); G.W. Anderson and S.M. Carroll,
%astro-ph/9711288; D. Comelli, M. Pietroni and A. Riotto, {\it
%Phys. Lett.} {\bf B571}, 115 (2003).

%\bibitem{diff}
% D.J. Holden and
%D. Wands, {\it Phys. Rev.} {\bf D61}, 043506 (2000); A.P. Billyard
%and A.A. Coley, {\it Phys. Rev.} {\bf D61}, 083503 (2000); N.
%Bartolo and M. Pietroni, {\it Phys. Rev.} {\bf D61}, 023518
%(2000);
 %M. Gasperini,
%{\it Phys. Rev.} {\bf D64}, 043510 (2001); A. Albrecht, C.P.
%Burges, F. Ravndal and C. Skordis, astro-ph/0107573; L.P.
%Chimento, A.S. Jakubi and D. Pavon, {\it Phys. Rev.} {\bf D62},
%063508 (2000); W. Zimdahl, D. Schwarz, A. Balakin and D. Pavon
%{\it Phys. Rev.} {\bf D64}, 063501 (2001); L.P. Chimento,  A.S.
%Jakubi, D. Pavon and W. Zimdahl, {\it Phys. Rev.} {\bf D67},083513
%(2003); A.A. Sen and S. Sen, {\it Mod. Phys. Lett.} {\bf A16},
%1303 (2001); G.R. Farrar and P.J.E. Peebles, astro-ph/0307316M;
%Axenides, K. Dimopoulos,  {\it JCAP} {\bf 0407}, 010,2004;  M.
%%Pietroni and L. Scarabello, {\it Phys.Rev.} {\bf D70}, 103526
%(2004).



%\bibitem{prob} D. Tocchini-Valentini and L Amendola, {\it Phys.
%Rev.}  {\bf D65}, 063508 (2002); L Amendola and D.
%Tocchini-Valentini, ibid., {\bf D66}, 043528 (2002); G.R. Farrar
%and P.J.E. Peebles, {\it Astrophys. J.} {\bf 604}, 1 (2004); M.B.
%Hoffman, astro-ph/0307350; U. Franca and R. Rosenfeld,
%astro-ph/0308149; astro-ph/0412413.

%\bibitem{longrange} See for example: R.D. Peccei, J. Sola, C.
%Wetterich, {\it Phys. Lett.} {\bf B195}, 183 (1987).

%\bibitem{carroll}
%S.M. Carroll, {\it Phys. Rev. Lett.} {\bf 81}, 3067 (1998).

%\bibitem{kolda}
%C. Kolda and D.H. Lyth, {\it Phys. Lett.} {\bf B458}, 197 (1999).
\bibitem{inflation}
%GThe Inflationary Universe: A Possible Solution to the Horizon and Flatness Problems
Alan H. Guth, 
Phys.Rev.D {\bf23} (1981) 347-356; Adv.Ser.Astrophys.Cosmol. 3 (1987) 139-148, 
A. Starobinsky, Phys.Lett. B {\bf91}, 99-102 (1980), 
E.W. Kolb and M.S. Turner,  ``The Early Universe'', Addison Wesley (1990); \\
A. Linde,  ``Particle Physics and Inflationary Cosmology'', Harwood, Chur,
Switzerland (1990); \\
A. Guth,  ``The Inflationary Universe'', Vintage, Random House (1998); \\
S. Dodelson,  ``Modern Cosmology'', Acad. Press (2003);\\
S. Weinberg,  ``Cosmology'', Oxford Univ. Press (2008), 
V. Mukhanov,  ``Physical Foundations of Cosmology'', Cambridge Univ. Press (2005). 
%%

\bibitem{Neutr-dark} P.Q. Hung, hep-ph/0010126; P.Q. Hung and H.
Pas, astro-ph/0311131; A. Singh, {\it Phys. Rev.} {\bf D52}, 6700
(1995); M. Blasone, A. Capolupo, S. Capozzielo, S. Carloni, G.
Vitiello, {\it Phys. Lett.} {\bf A323}, 182 (2004); X. Zhang,
hep-ph/0410292; X. Bi, B. Feng, H. Li and X. Zhang,
hep-ph/0412002; A.W. Brookfield, C.van de Bruck, D.F. Mota, D.
 Tocchini-Valentini, astro-ph/0503349; V. Barger, P. Huber and D.
 Marfatia, hep-ph/0502196; M. Cirelli, M.C. Gonzalez-Garcia and
 C. Pena-Garay, hep-ph/0503028; X.-f. Zhang, H. Li, Y.-S. Piao and
 X . Zhang, astro-ph/0501652.

 \bibitem{FNW-1} R. Fardon, A. E. Nelson, N. Weiner {\it JCAP}
 {\bf 0410}, 005 (2004).

 \bibitem{FNW-2} R.D. Peccei, {\it Phys. Rev.} {\bf D71}, 023527 (2005).

 \bibitem{GK1}
E.I. Guendelman and A.B. Kaganovich, {\it Phys. Rev.} {\bf D53},
7020 (1996); {\it Mod. Phys. Lett.} {\bf A12}, 2421 (1997); {\it
Phys. Rev.} {\bf D55}, 5970 (1997);  {\it Mod. Phys. Lett.} {\bf
A12}, 2421 (1997); {\it Phys. Rev.} {\bf D55}, 5970 (1997); ibid.
{\bf D57}, 7200 (1998); {\it Mod. Phys. Lett.} {\bf A13}, 1583
(1998).

\bibitem{GK2}
E.I. Guendelman and A.B. Kaganovich, {\it Phys. Rev.} {\bf D56},
3548 (1997).

\bibitem{GK3} E.I. Guendelman and A.B. Kaganovich, {\bf Phys.
Rev.}
{\bf D60}, 065004 (1999).

\bibitem{G1}
 E.I. Guendelman, {\it Mod.
Phys. Lett.} {\bf A14}, 1043 (1999);
 {\it Class. Quant. Grav.} {\bf 17}, 361 (2000);
gr-qc/9906025; {\it Mod. Phys. Lett.} {\bf A14}, 1397 (1999);
gr-qc/9901067; hep-th/0106085; {\it Found. Phys.} {\bf 31}, 1019
(2001);
\bibitem{G2}
E.I. Guendelman, {\it Phys. Lett.} {\bf B412}, 42 (1997); E.I.
Guendelman, gr-qc/0303048; E.I. Guendelman and E. Spallucci,
hep-th/0311102.


\bibitem{tfr}
 E.I.
Guendelman and O. Katz, {\it Class. Quant. Grav.}, {\bf 20}, 1715
(2003).
\bibitem{conSergio1}
Sergio del Campo, Eduardo I. Guendelman, Ramon Herrera, Pedro Labraña.
Emerging Universe from Scale Invariance,  JCAP 06 (2010) 026 • e-Print: 1006.5734 [astro-ph.CO]
 \bibitem{conSergio2}
 Sergio del Campo, Eduardo I. Guendelman, Alexander B. Kaganovich, Ramon Herrera, Pedro Labrana, Emergent Universe from Scale Invariant Two Measures Theory, Phys.Lett.B 699 (2011) 211-216 • e-Print: 1105.0651 [astro-ph.CO]
 \bibitem{EMILSVETLANA} Eduardo Guendelman, Ramón Herrera, Pedro Labrana, Emil Nissimov, Svetlana Pacheva.
 Emergent Cosmology, Inflation and Dark Energy
  Gen.Rel.Grav. 47 (2015) 2, 10 • e-Print: 1408.5344 [gr-qc]
\bibitem{curvaton1}  Eduardo I. Guendelman, Ramón Herrera,
Curvaton reheating mechanism in a scale invariant two measures theory
  Gen.Rel.Grav. 48 (2016) 1, 3 • e-Print: 1511.08645 [gr-qc]; S.~del Campo and R.~Herrera,
%``Curvaton field and intermediate inflationary Universe model,''
Phys. Rev. D \textbf{76}, 103503 (2007)
doi:10.1103/PhysRevD.76.103503
[arXiv:0710.5524 [astro-ph]]; S.~del Campo, R.~Herrera, J.~Saavedra, C.~Campuzano and E.~Rojas,
%``Curvaton reheating in logamediate inflationary model,''
Phys. Rev. D \textbf{80}, 123531 (2009)
doi:10.1103/PhysRevD.80.123531
[arXiv:0912.4721 [astro-ph.CO]].
\bibitem{curvaton2}
Eduardo I. Guendelman, Ramon Herrera, Pedro Labrana, 
Instant preheating in a scale invariant two measures theory
 Phys.Rev.D 103 (2021) 123515 • e-Print: 2005.14151 [gr-qc]
 \bibitem{H0} Eduardo Guendelman, Ramón Herrera, David Benisty,
 Unifying inflation with early and late dark energy with multiple fields: Spontaneously broken scale invariant two measures theory
, Phys.Rev.D 105 (2022) 12, 124035 • e-Print: 2201.06470 [gr-qc]
 \bibitem{H0DM} Eduardo Guendelman, Ramon Herrera, Unification: Emergent universe followed by inflation and dark epochs from multi-field theory, Annals Phys. 460 (2024) 169566 • e-Print: 2301.10274 [gr-qc]
\bibitem{Guendelman:2006ji}
E.~I.~Guendelman and A.~B.~Kaganovich,
%``Exotic low density fermion states in the two measures field theory: Neutrino dark energy,''
Int. J. Mod. Phys. A \textbf{21}, 4373-4406 (2006)
doi:10.1142/S0217751X06032538
[arXiv:gr-qc/0603070 [gr-qc]].

\bibitem{reh}
L. Abbott, E. Farhi, and M. B. Wise, Particle Production in the New Inflationary Cosmology, Phys.Lett. B {\bf117}, 29 (1982);
A. Dolgov and A. D. Linde, Baryon Asymmetry in Inflationary Universe, Phys.Lett. B {\bf116},329 (1982).



\bibitem{Prod} L. Surhone, M. Timplendon and S. Marseken, {\em ``Wright Omega Function: Mathematics, Lambert W Function,
Continuous Function, Analytic Function, Differential Equation,
Separation or Variables"}, Betascript Publishing (2010). 

\bibitem{p1} X.~Chen, M.~X.~Huang, S.~Kachru and G.~Shiu,
\textsl{JCAP} {\bf 0701} (2007) 002; B.~A.~Bassett, S.~Tsujikawa and D.~Wands,
%``Inflation dynamics and reheating,''
Rev. Mod. Phys. \textbf{78},  (2006) 537-589.

\bibitem{Planck:2018vyg}
N.~Aghanim \textit{et al.} [Planck],
%``Planck 2018 results. VI. Cosmological parameters,''
Astron. Astrophys. \textbf{641}, A6 (2020)
[erratum: Astron. Astrophys. \textbf{652}, C4 (2021)]
doi:10.1051/0004-6361/201833910
[arXiv:1807.06209 [astro-ph.CO]].

\bibitem{Gasperini}
F.W. Hehl, P. v. d. Heyde and G. Kerlick, {\it Reviews of Modern
Physics} {\bf 48}, 393 (1976); V. de Sabbata and M. Gasperini,
{\it Introduction to Gravitation},
 (World Scientific Publishing, Singapure, 1985).





\end{thebibliography}
\end{document}